# Modulation instability in nonlinear metamaterials induced by cubic-quintic nonlinearities and higher order dispersive effects

**Manirupa Saha and Amarendra K. Sarma***
Department of Physics, Indian Institute of Technology Guwahati, Guwahati-781039, Assam, India.
*E-mail: aksarma@iitg.ernet.in

**Abstract:**
We have investigated modulation instability (MI) in metamaterials (MM) with both cubic and quintic nonlinearities, based on a model appropriate for pulse propagation in MM with cubic-quintic nonlinearities and higher order dispersive effects. We have included loss into account in our analysis and found that loss distorts the sidebands of the MI gain spectrum. We find that the combined effect of cubic-quintic nonlinearity increases the MI gain. The role of higher order nonlinear dispersive effects on MI has also been discussed.

*Key words:* Modulation instability; Negative index material; Metamaterials; Nonlinear Schrodinger Equation.

## 1. INTRODUCTION

Modulation instability (MI) is a fundamental and ubiquitous process that appears in most nonlinear systems in nature [1-7]. It occurs as a result of interplay between the nonlinearity and dispersion in the time domain or diffraction in the spatial domain. In the MI process, weak perturbations imposed on a continuous wave (cw) state grow exponentially due to the interplay between nonlinear and dispersive effects. As time goes on, the modulation increases and the cw wave breaks into a periodic pulse train. In the Fourier spectrum MI manifests itself as a pair of sideband components around the carrier wave component [8]. MI phenomena have been studied quite extensively in many areas of science and engineering. In optics, the interest in MI stems from its possible applications and relevance in ultrafast pulse generation [9], supercontinuum generation [10], four-wave mixing [11], Bragg gratings [12], quadratic media [7], optical fiber [1,5], parametric oscillators [13] etc. In general, MI occurs in the anomalous group velocity dispersion (GVD) regime for a focusing nonlinearity [1]. MI is possible in the normal GVD regime also but it is limited to some special cases [14-19]. Recently, MI study has received lots of attention in the context of metamaterials (MM), specifically in the so-called negative-index metamaterials (NIM) [2]. In this work we use the terms MM and NIM in the same sense. It may be noted that MM are man-made artificial structures which exhibit uncommon properties, so far unavailable in naturally occurring materials, such as reversal of Snell's law, reversal of Doppler effect and Cherenkov effect, etc. [20]. One key feature of MM is that they show negative refractive index, for which they are often referred to as negative index material (NIM), owing to their simultaneous negative electric permittivity and negative magnetic permeability. The first possibility of NIM was theoretically introduced by Veselago in 1968 [21]. Because of the recent experimental realization of NIM at infrared and optical frequencies [22-24] MM or NIM research is getting a new dimension. Various potential applications of MM have been proposed and studied [25-27]. As stated earlier, MI study in the context of MM has turned out to be an extremely exciting field of research. One primary reason among many others is related to the fact that MI could be considered as a precursor to soliton formation and MM may offer new ways to generate and control bright and dark solitons [28-31]. In this work we report a MI analysis in a nonlinear material induced by cubic-quintic nonlinearities and higher order dispersive effects. The analysis is based on a mathematical model derived in a similar approach followed by authors in Refs. [2, 32-34]. It is worthwhile to note that some conventional materials exhibit higher order nonlinear susceptibilities even at moderate pulse intensity. For example, $CdS_xSe_{1-x}$-doped glass possesses a considerable amount of fifth-order susceptibility $\chi^{(5)}$ [35]. Again it has been demonstrated that transparent glass in intense femtosecond pulse at 620 nm shows significant nonlinear effects due to $\chi^{(5)}$ [36]. A recent experiment even shows that materials such as chalcogenide glass exhibit not only third and fifth order nonlinearities but even seventh-order nonlinearity [37]. Motivated by these developments, in this work, we are investigating modulation instability induced by cubic-quintic nonlinearities. We have derived an appropriate mathematical model starting with the well-known Maxwell's equations to take into account the effect of higher order dispersion upto fourth order along with nonlinear susceptibilities owing to $\chi^{(3)}$ and $\chi^{(5)}$. The model successfully reproduces previously proposed similar models under appropriate approximations from our newly derived propagation equation. It should be noted that so far many models related to the pulse propagation in nonlinear NIM have been proposed, including the interaction of ultrashort pulses with MMs [32-34]. Recently, a generalized nonlinear Schrodinger equation (NLSE) suitable for few-cycle pulse propagation in the MM's with delayed Raman response is reported [38]. In this work, however, we would mainly be concerned with effects arising owing to cubic-quintic nonlinearities in MMs. Finally, we need to note that the modulation instability in conventional nonlinear materials with both cubic-quintic nonlinearities have been discussed by many authors in various contexts [39,40]. Again, the study of MI is not limited to a single channel negative index material, it is extended to the case of coupled channels as well [41], due to many possible applications and novel phenomena [42]. The difference between the MI in ordinary or conventional materials and negative index materials fundamentally occurs owing to the fact that NIMs are artificial materials exhibiting both dispersive magnetic permeability and electric permittivity. One can tune the optical response of MMs through structure design, which may not be the case with the so called conventional materials. Even the nonlinearity of MMs could be engineered with judicious choice or design of the constituent metamolecules [43]. MI in MM in the presence of cubic-quintic nonlinearities and higher order dispersive effects has a better controllability compared to the common nonlinear materials. Another important feature of MI in MM is that it may be triggered in both the normal and the anomalous dispersion regimes with judicious choice of the parameters.

## 2. THEORETICAL MODEL

The pulse propagation equation in nonlinear MM with both $\chi^{(3)}$ and $\chi^{(5)}$ nonlinearities are obtained by generalising the derivation of the so called Nonlinear Schrodinger equation (NLSE) based on previous works [32-34]. It is worthwhile to mention that one of the most fundamental differences between an ordinary material and an MM is that MM exhibits strong dispersive behaviour in electric permittivity and magnetic permeability, while, in ordinary materials, only one of them is dominant at a time. Now, starting with Maxwell's equations and adopting a procedure similar to that of Ref. 33, we obtain the following one dimensional pulse propagation equation in a nonlinear negative index metamaterial:

$$\frac{\partial A}{\partial \xi} = -\frac{i\beta_2}{2}\frac{\partial^2 A}{\partial \tau^2} + \frac{\beta_3}{6}\frac{\partial^3 A}{\partial \tau^3} + \frac{i\beta_4}{24}\frac{\partial^4 A}{\partial \tau^4} + i\sigma_0\left(|A|^2 A + iS_1\frac{\partial}{\partial \tau}\left(|A|^2 A\right) - S_2\frac{\partial^2}{\partial \tau^2}\left(|A|^2 A\right)\right)$$
$$+i\eta_0'|A|^4 A - \eta_0 S_3\frac{\partial}{\partial \tau}\left(|A|^4 A\right) \qquad (1)$$

In deriving the generalized NLSE, we assume the nonlinear electric polarization ($P_{NL}$) as $\mathbf{P}_{NL} = \varepsilon_{NL}\mathbf{E} = \varepsilon_0 \chi^{(3)}|E|^2 \mathbf{E} + \varepsilon_0 \chi^{(5)}|E|^4 \mathbf{E}$; where $\mathbf{E}$ is the electric filed, $\varepsilon_{NL}$ is the nonlinear electric permittivity and $\chi^{(n)}$ is the n-th order electric susceptibility. The dielectric permittivity (ε) and magnetic permeability (µ) are dispersive in MMs and their frequency dispersion is given by the lossy Drude model [44]:

$$\varepsilon(\tilde{\omega}) = \varepsilon_0\left[1 - 1/\left(\tilde{\omega}(\tilde{\omega} + i\tilde{\gamma}_e)\right)\right]; \mu(\tilde{\omega}) = \mu_0\left[1 - \tilde{\omega}_{pm}^2/\left(\tilde{\omega}(\tilde{\omega} + i\tilde{\gamma}_m)\right)\right]$$
(2)

Here $\tilde{\omega} = \omega/\omega_{pe}$, $\tilde{\omega}_{pm} = \omega_{pm}/\omega_{pe}$ with $\omega_{pe}$ and $\omega_{pm}$ are the respective electric and magnetic plasma frequencies. $\tilde{\gamma}_e = \gamma_e/\omega_{pe}$ and $\tilde{\gamma}_m = \gamma_m/\omega_{pe}$ are the electric and magnetic loss respectively normalized with respect to the electric plasma frequency. $\varepsilon_0$ and $\mu_0$ are respectively the free space electric permittivity and magnetic permeability. It should be noted that loss is an extremely important and relevant issue in MMs as shown by the recent work of Stockman [45]. In this work we are taking $\tilde{\gamma}_e \sim \tilde{\gamma}_m \sim 0.01$, a value two orders of magnitude greater than that of the one taken by D'Aguanno et al. [44]. In Eq. (2), A is the slowly varying envelope of the pulse, propagating along the $\xi$ direction. $\beta_2, \beta_3$ and $\beta_4$ are the group velocity dispersion(GVD), the third and the fourth order dispersion parameters respectively, while $\sigma_0$ and $\eta_0'$ are the cubic and quintic nonlinear coefficients respectively. $S_1$ is the so-called self-steepening (SS) parameter due to cubic nonlinear polarization while $S_3$ is the SS parameter due to quintic nonlinear polarization. On the other hand, $S_2$ is the second order nonlinear dispersive coefficient. The above mentioned parameters are defined as follows:

$$\beta_2 = \delta_2/2k_0 - 1/k_0 V^2; \beta_3 = \delta_3 - 3\beta_2/k_0 V; \beta_4 = \delta_4 - 3\beta_2^2/k_0 - 4\delta_3/k_0 V;$$
$$S_1 = 1/\omega_0 - i\sigma_1/\sigma_0 - 1/k_0 V; S_2 = -\sigma_2/\sigma_0 - i\sigma_1/\sigma_0\omega_0 - \beta/4k_0 - 1/\omega_0 k_0 V + i\sigma_1/k_0 V\sigma_0; \qquad (3)$$
$$S_3 = 1/\omega_0 - i\eta_1/\eta_0 - 1/k_0 V,$$

with

$$\delta_m = \frac{m!}{2k_0}\sum_{l=0}^{m}\frac{1}{l!m-l!}\frac{\partial^l(\omega\varepsilon)}{\partial\omega^l}\bigg|_{\omega=\omega_0}\frac{\partial^{m-l}(\omega\mu)}{\partial\omega^{m-l}}\bigg|_{\omega=\omega_0}; V = 2k_0/\omega_0(\varepsilon(\omega)\gamma + \mu(\omega)\alpha);$$
$$\sigma_m = m!\omega_0\varepsilon_0\chi^{(3)}F_m/2k_0; \eta_m = m!\omega_0\varepsilon_0\chi_E^{(5)}F_m/2k_0, \qquad (4)$$

where $F_m = \frac{i^m}{m!} \frac{\partial^m (\omega \mu)}{\partial \omega^m}\bigg|_{\omega=\omega_0}$, $k_0 = n\omega_0/c$, $\alpha = \partial\{\omega\varepsilon(\omega)\}/\partial\omega|_{\omega=\omega_0}$ and $\gamma = \partial\{\omega\mu(\omega)\}/\partial\omega|_{\omega=\omega_0}$

The non-SVEA correction terms to first order is approximated as:

$$\frac{\partial^2 A}{\partial \xi^2} \approx -\frac{\beta_2^2}{4}\frac{\partial^4 A}{\partial \tau^4} - \sigma_0^2 |A|^4 A + \frac{\sigma_0 \beta_2}{2}\frac{\partial^2}{\partial \tau^2}|A|^2 A \tag{5}$$

$$\frac{\partial^2 A}{\partial \xi \partial \tau} \approx -\frac{i\beta_2}{2}\frac{\partial^3 A}{\partial \tau^3} + \frac{\delta_3}{6}\frac{\partial^4 A}{\partial \tau^4} + i\sigma_0 \frac{\partial}{\partial \tau}|A|^2 A + i\eta_0 \frac{\partial}{\partial \tau}|A|^4 A - \frac{\sigma_0}{\omega_0}\frac{\partial^2}{\partial \tau^2}|A|^2 A + i\sigma_1 \frac{\partial^2}{\partial \tau^2}|A|^2 A$$

(6)

It is worthwhile to mention that the fourth order dispersion parameter $\beta_4$ contains one extra term while the second order nonlinear dispersive parameter $S_2$ has two additional terms as compared to the ones considered in an earlier work [46]. This difference can be attributed to the fact that, in this work, we are considering both the cubic and the quintic nonlinearities. In the expression for $\beta_4$, we are getting an additional term due to the third order dispersion (TOD) effect.

For the purpose of simplification it is convenient to write Eq. (2) in the normalized form. We assume the normalized variables as [1]:

$$Z = \xi/L_D, T = \tau/T_0, A = \sqrt{P_0} U \quad u = NU \tag{7}$$

Here $L_D = T_0^2/|\beta_2|$ is the so-called second order dispersion length with $T_0$ representing the pulse width. We can similarly define the third and the fourth order dispersion length as $L_D' = T_0^3/\beta_3$ and $L_D'' = T_0^4/\beta_4$ respectively. N is termed as the order of the soliton, defined as $N^2 = L_D/L_{NL}$ with $L_{NL} = 1/\sigma_0 P_0$, the so-called nonlinear length.

Eq. (2) can thus be written in the dimensionless units as follows:

$$\frac{\partial u}{\partial Z} = -\frac{ik_2}{2}\frac{\partial^2 u}{\partial T^2} + \frac{k_3}{6}\frac{\partial^3 u}{\partial T^3} + \frac{ik_4}{24}\frac{\partial^4 u}{\partial T^4} + i|u|^2 u + ip|u|^4 u - s_1 \frac{\partial}{\partial T}(|u|^2 u) - is_2 \frac{\partial^2}{\partial T^2}(|u|^2 u) - s_3 p' \frac{\partial}{\partial T}(|u|^2 u) \tag{8}$$

Here $k_2 = \text{sgn}(\beta_2)$, $k_3 = L_D/L_D'$ and $k_4 = L_D/L_D''$. $s_1 = |S_1|/T_0$ and $s_2 = |S_2|/T_0^2$ are the normalized self-steepening (SS) parameter and second order normalized nonlinear dispersion parameter respectively, while $s_3 = |S_3|/T_0$ is the normalized SS parameter due to quintic nonlinearity. $p = \eta_0'/\sigma_0^2 L_D - 1/2k_0 L_D$ is the normalized quintic nonlinear parameter and $p' = \eta_0'/\sigma_0^2 L_D$.

## 3. MODULATION INSTABILITY ANALYSIS

We would now, on the basis of Eq. (8), investigate the MI by using the standard linear stability analysis [1]. Eq. (8) has a steady state solution given by $u = \sqrt{P}\exp[i(P+P^2 p)Z]$, where P is the normalized optical power. We introduce perturbation $a(Z,T)$, such that $a(Z,T) \ll \sqrt{P}$, together with the steady state solution to Eq. (8) and linearize in $a(Z,T)$ to obtain:

$$\frac{\partial a}{\partial Z} = -\left(\frac{k_2}{2} + 2s_2 P\right)\frac{\partial^2 a}{\partial T^2} + \frac{k_3}{6}\frac{\partial^3 a}{\partial T^3} + \frac{k_4}{24}\frac{\partial^4 a}{\partial T^4} + iP(a+a^*) + 2ipP^2(a+a^*) - (2s_1 P + 3s_3 p' P^2)\frac{\partial a}{\partial T}$$
$$+ (s_1 P + 2s_3 p' P^2)\frac{\partial a^*}{\partial T} - is_2 P \frac{\partial^2 a^*}{\partial T^2}$$

(9)

Now, writing $a = a_1 e^{i(KZ-\Omega T)} + a_2 e^{-i(KZ-\Omega T)}$, where K and $\Omega$ are the normalized wavenumber and the frequency of perturbation respectively, from Eq. (9) we obtain the following dispersion relation:

$$K = \frac{k_3}{6}\Omega^3 + (2s_1 P + 3s_3 p' P^2) \pm \sqrt{\Omega^4 M + \Omega^2 N}.$$ (10)

Here, $M = \left(\frac{k_2}{2} + 2s_2 P + \frac{k_4}{24}\Omega^2\right)^2 - s_2^2 P^2$ and

$$N = \left[2\left(\frac{k_2}{2} + 2s_2 P + \frac{k_4}{24}\Omega^2\right)(P + 2P^2 p) - 2s_2 P(P + 2P^2 p) + (s_1 P + 2s_3 p' P^2)^2\right]$$

The steady-state solution becomes unstable whenever K has an imaginary part since the perturbation then grows exponentially. One can easily see that for the occurrence of MI we must have $\Omega^2 M + N < 0$. Under this condition, the gain spectrum $g(\Omega)$ of the MI could be expressed as:

$$g(\Omega) = 2\,\text{Im}(K) = 2|\Omega|\sqrt{N + \Omega^2 M}$$ (11)

The MI gain spectrum depends on the parameters M and N, which in turn depend on parameters like P, $s_2$, $s_3$, p etc. Hence, MI in MM in the presence of cubic-quintic nonlinearities and higher order dispersive effects has a better controllability compared to the common nonlinear materials. In Fig.1 we show the variation of the MI gain with normalized perturbation frequency for different values of the normalized frequency $\omega/\omega_{pe}$ with $\gamma_e = 0.01$. We observe that the MI gain amplitude as well as the gain bandwidth is reduced with increase of the normalized frequency $\omega/\omega_{pe}$. The appearance of distortions in the sidebands could be attributed to the presence of linear loss. The influence of loss on modulation instability is briefly discussed in a separate section.

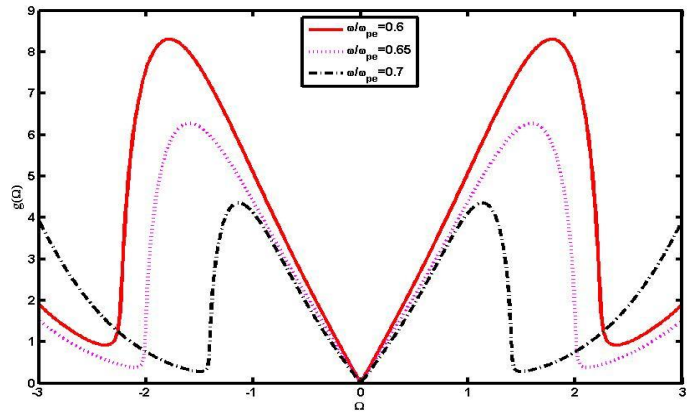

Fig.1. (color online) MI gain as a function of perturbation frequency $\Omega$ for different values of $\omega/\omega_{pe}$ with $\gamma_e = 0.01$.

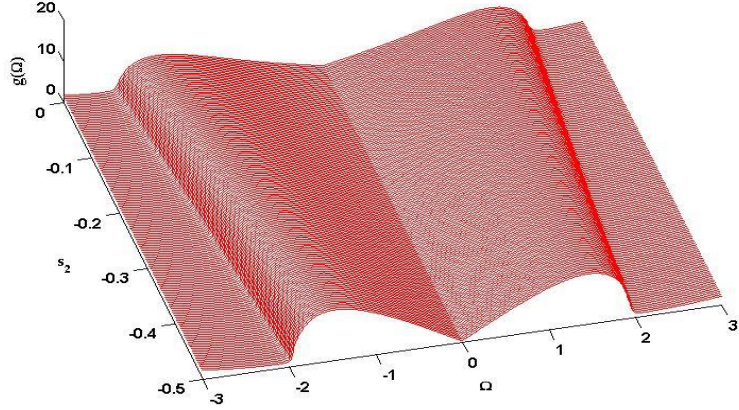

Fig.2. (color online) MI gain spectrum with normalized second order self-steepening parameter $s_2$ and perturbation frequency $\Omega$ for P=2 and $s_3$=0 in the anomalous dispersion region.

The role of the SS parameter $s_1$ on MI has been studied earlier and it was shown that MI gain decreases and the gain band width also shrinks owing to self-steepening effect and it does not depend on the sign of the SS parameter $s_1$[47]. In Fig. 2 we depict the role of the second order nonlinear dispersion parameter $s_2$ on MI gain with P=2, $s_1 = -0.168$ and $s_3 = 0$ for a pulse width $T_0$=10 fs and $\omega_0 / \omega_{pe} = 0.6$. We notice that as $s_2$ parameter increases the MI gain decreases while the bandwidth enlarges. We have also investigated the role of the nonlinear parameter p, which results due to the combined effect of cubic-quintic nonlinearities, on the MI gain spectrum. From Fig. 3(a) we observe that MI gain as well as the gain band width increases with the increase of the nonlinear parameter p. This could also be seen clearly from Fig. 3(b)

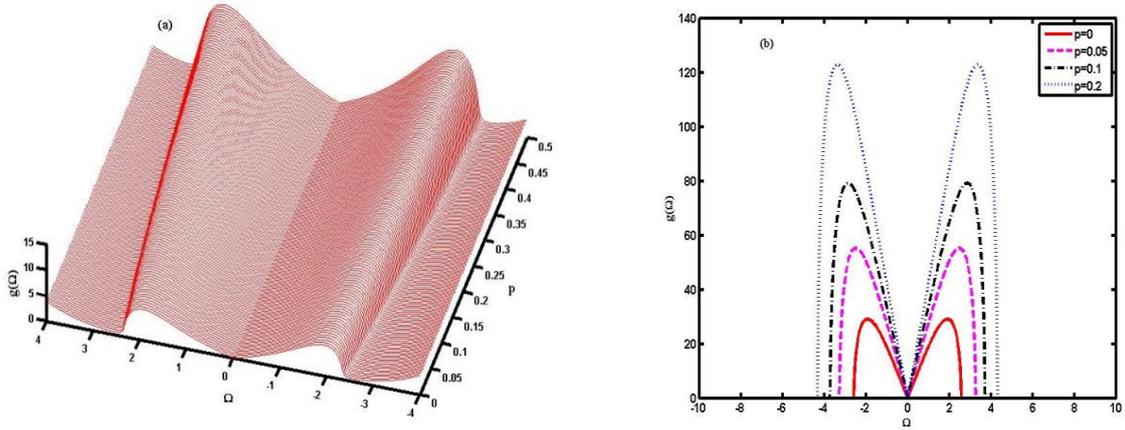

Fig.3 (color online) (a) MI gain as a function of perturbation frequency $\Omega$ and higher order nonlinear parameter p with fixed value of $s_2$. (b) MI gain as a function of higher order nonlinear parameter p.

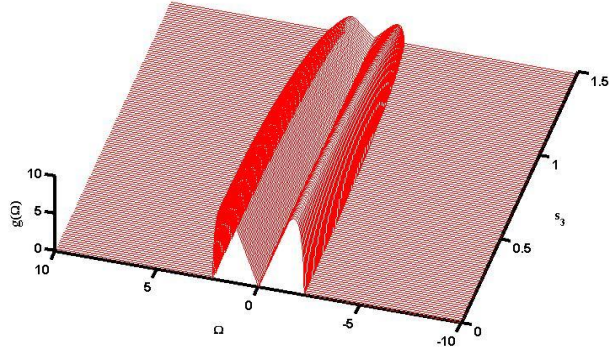

Fig.4. (color online) MI gain as a function of perturbation frequency $\Omega$ and higher order self-steepening parameter $s_3$ with fixed value of $s_2$.

The effect of $s_3$, which arises due to quintic nonlinearity, on MI gain is shown in Fig.4. It could be seen that $s_3$ suppresses the MI gain and reduces the bandwidth, a feature exhibited also by the self-steepening parameter $s_1$ [47]. We find that MI gain is also independent of the sign of $s_3$.

## 4. INFLUENCE OF LOSS ON MODULATION INSTABILITY

Linear loss is an important and very relevant issue in metamaterials [44,45,48]. At optical wavelengths MM suffer from high dissipative losses due to the metallic nature of their constituent metamolecules. Hence there is a great thrust in research in overcoming the loss restrictions in MM. However this issue of loss is very complex in the so called MM or NIM. For example, based on a causality study, it has been shown theoretically that if the losses in MM are eliminated or significantly reduced by any means, including the compensation by gain media, then the negative refraction will disappear [45]. This clearly shows that one has to be careful in addressing any issue related to loss in MM. As regards, MI is concerned, the impact of loss on MI have been studied in conventional materials, such as an optical fiber, is studied by some authors [49-50] . They have shown that in the presence of loss, the cw background decays during propagation, and the associated gain spectrum changes. It has been shown that the effectiveness of MI processes in high index optical fibers is not merely influenced by the strength of the nonlinearity, but is also strongly determined by the linear attenuation of waves in the fiber material. In those high-index glass fibers, this attenuation acts as a strong perturbation, causing a frequency drift of the MI sidebands. However, techniques to suppress this frequency drift is completely developed. In contrast to the ordinary materials, in MM, the impact of loss on MI is generally avoided , may be for the reasons mentioned above. The issue of loss is generally included in the mathematical model through the so called Drude model. In order to see the role of loss on MI, in Fig. 5, we depict the MI gain $g(\Omega)$ as a function of the perturbation frequency for different loss parameters with normalized peak power at P=2.

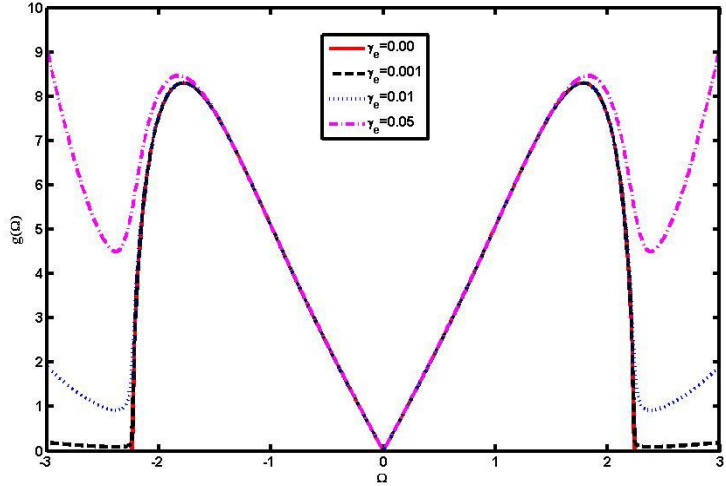

Fig.5: Modulation Instability gain Vs perturbation frequency with different loss parameter with peak power P=2.

It can be observed from Fig. 5 that when we consider the Drude model without any loss we get two prominent sidebands. But as loss is incorporated, with increase in the value of the loss coefficient, the peak of the MI gain remains nearly the same but the sidebands get distorted. However at $\tilde{\gamma}_e \sim \tilde{\gamma}_m \sim 0.01$, a value two orders of magnitude greater than that of the one taken by D'Aguanno et al. [44], peak of the side bands are distinct and distortions in the sideband could be fairly neglected. Finally, we would like to conclude that our study of the impact of loss on MI in nonlinear metamaterials is not a detailed or rigorous one. However, this study may enhance some interest in this very critical issue related to MMs.

## 5. CONCLUSIONS

To conclude, we have presented a modulation analysis based on a model appropriate for pulse propagation in MM with cubic-quintic nonlinearities and higher order dispersive effects. We have included loss into account in our analysis and found that loss distorts the sidebands of the MI gain spectrum. However, in this work we are taking the loss parameters to be $\tilde{\gamma}_e \approx \tilde{\gamma}_m = 0.01$, a value two orders of magnitude greater than that of the one taken by D'Aguanno et al. [44], at which the peak of the side bands are distinct and distortions in the sideband could be fairly neglected. We find that the combined effect of cubic-quintic nonlinearity increases the MI gain. The role of higher order nonlinear dispersive effects on MI has been also discussed. It is shown that self-steepening parameter $s_3$, which arises due to $\chi^{(5)}$ nonlinear polarisation, exhibits similar MI features with that of the self-steepening parameter $s_1$. It is clear that with so many controllable parameters, MM or NIM embedded in cubic quintic nonlinear medium may provide us more enriching features to manipulate modulation instability and there by soliton generation or control in metamaterials.